\documentclass[%
 aip,
 jmp,%
 amsmath,amssymb,
reprint,%
]{revtex4-1}

\usepackage{graphicx}
\usepackage{dcolumn}
\usepackage{bm}
\usepackage{subfigure} 

\begin{document}


\title[Nucleation and growth by diffusion]{Nucleation and growth by diffusion under Ostwald-Freundlich boundary condition
}

\author{Masao Iwamatsu}
\email{iwamatsu@ph.ns.tcu.ac.jp}
\affiliation{ 
Department of Physics, Faculty of Liberal Arts and Science, Tokyo City University, Setagaya-ku, Tokyo 158-8557, JAPAN
}%

\date{\today}

\begin{abstract}
The critical radius of a nucleus grown by diffusion in a solution is studied thermodynamically as well as kinetically.  The thermodynamic growth equation called Zeldovich equation of classical nucleation theory (CNT) and the kinetic diffusional growth equation combined with the Ostwald-Freundlich boundary condition lead to the same critical radius.  However, it should be pointed out that the diffusional equation may lead to a kinetic critical radius that is different from the thermodynamic critical radius, thus indicating the possibility of kinetically controlling the critical radius of a nucleus. 

\end{abstract}

\maketitle

Nucleation and growth are basic phenomena that play vital roles in the processing of various materials across industries~\cite{Kelton2010}.  In particular, the growth of various fine particles such as semiconductor quantum dots~\cite{Gorshkov2010}, bio-minerals~\cite{Meldrum2008} and other molecular crystals in solution~\cite{Sear2012} has attracted considerable interest recently.  In those materials, the material transport by diffusion plays fundamental role. This growth problem by diffusion has been studied, for example, in the precipitation from　solution~\cite{Zener1949}, the liquid droplet nucleation from supersaturated vapor~\cite{Frank1950}, the vapor bubble nucleation from supersaturated ~\cite{Epstein1950} solution, and the colloidal particle formation~\cite{LaMer1950,Reiss1951} from solution. However, a more complex scenario has been suggested recently, for example, to explain the mono-dispersed colloidal particles~\cite{Privman1999}, where primary particles produced by the nucleation and subsequent growth will aggregate to form mono-dispersed secondary particles~\cite{Robb2008}. In most of previous studies, however, either nucleation or growth is studied separately~\cite{Robb2008,Kuchma2009a, Grinin2011} and a fully consistent formulation of nucleation and growth is missing. 

In fact, nucleation and growth can be studied on the same footing using the general formulation using master equation called Becker-D\"oring equation~\cite{Kalikmanov2013} or using Fokker-Planck equation~\cite{Kalikmanov2013}. Although, they can be used to study complex nucleation in general, the result is mostly numerical~\cite{Wilemski1995,Wyslouzil1995,Kathmann2004} in the abstract phase space and cannot give a clear picture of nucleation by fluctuation and growth by diffusion in real space.

In this communication, we will look at the nucleation and growth by diffusion~\cite{Zeldovich1943,Zener1949,Reiss1951} of the post-critical nucleus.  We closely follow the discussions of Slezov~\cite{Slezov2009} and Peter~\cite{Peter2011}, and observe the relationship between the Ostwald-Freundlich boundary condition and the critical radius of a nucleus, which is usually assumed to be determined solely by the thermodynamical growth equation~\cite{Iwamatsu2012}.

We concentrate on the one-component system and consider nucleation from solution within the context of the classical nucleation theory (CNT)~\cite{Kelton2010}.  We start from the fundamental kinetic equation for the evolution of the cluster distribution function $f(n,t)$, which represents the number of clusters consisting of $n$ monomers.  This is written as
\begin{eqnarray}
\frac{\partial f\left(n,t\right)}{\partial t} &=& w_{n-1,n}^{(+)}f\left(n-1,t\right)-w_{n,n-1}^{(-)}f\left(n,t\right) \nonumber
\\
&&+w_{n+1,n}^{(-)}f\left(n+1,t\right)-w_{n,n+1}^{(+)}f\left(n,t\right),
\label{eq:s1}
\end{eqnarray}
where $w_{n-1,n}^{(+)}$ is the rate of attachment of a monomer onto a cluster consisting of $n-1$ monomers.  Similarly, $w_{n,n-1}^{(-)}$ is the rate of detachment of a monomer from a cluster that consists of $n$ monomers.  By introducing the virtual state, that is equilibrium with the cluster consisting of $n$ monomers~\cite{Slezov1998}, this master equation is usually transformed into the well-known Fokker-Planck equation that describes probability flow in a space of size ($n$) :
\begin{equation}
\frac{\partial f\left(n,t\right)}{\partial t} = \frac{\partial}{\partial n}\left[w\left(n\right)\left\{\beta\frac{\partial \Delta G\left(n\right)}{\partial n}f\left(n,t\right)+\frac{\partial f\left(n,t\right)}{\partial n}\right\} \right],
\label{eq:s2}
\end{equation}
where $w\left(n\right)$ denotes continuum version of the rate of attachment $w_{n,n+1}^{(+)}$, $\beta=1/k_{\rm B}T$ denotes the usual inverse temperature, and $\Delta G\left(n\right)$ denotes the free energy of a cluster consisting of $n$ monomers.  Within the CNT, it is given by
\begin{equation}
\Delta G\left(n\right)=-n\Delta\mu_{0}+\gamma\phi n^{2/3},
\label{eq:s3}
\end{equation}
where 
\begin{equation}
\Delta \mu_{0}=k_{\rm B}T\ln\left(c_{0}/c_{\rm sat}\right),
\label{eq:s4}
\end{equation}
is the chemical potential of a solution with the concentration $c_{0}$ that is higher than the saturation concentration $c_{\rm sat}$ ($c_{0}>c_{\rm sat}$).  Therefore, the solution is supersaturated and $\Delta \mu_{0}>0$. $\gamma$ is the surface tension of the cluster (nucleus), and $\phi$ is a shape factor that connects the radius of the cluster, $R$, to the number of monomers, $n$, in the cluster through $4\pi R^{2}=\phi n^{2/3}$ for spherical clusters. 

Because the second term on the right-hand side of Eq.~(\ref{eq:s2}) is similar in form to the usual diffusion equation in real space, this term describes diffusion in $n$ space and is called the diffusion term. While the first term describes cluster growth through
\begin{equation}
\frac{dn}{dt}=-\beta w\left(n\right)\frac{\partial \Delta G\left(n\right)}{\partial n}=-\beta w\left(n\right) \mu\left(n\right) 
\label{eq:s5}
\end{equation}
and is called the drift term, wherein we have used the chemical potential $\mu\left(n\right)$ of the cluster with $n$ monomers defined as
\begin{equation}
\mu\left(n\right) = \frac{\partial \Delta G}{\partial n}= -\Delta \mu_{0}+\frac{2\gamma\phi}{3}n^{-1/3}.
\label{eq:s6}
\end{equation}
Eq.~(\ref{eq:s5}) is also know as the Zeldovich equation~\cite{Zeldovich1943,Kalikmanov2013}; it describes nucleus growth in terms of thermodynamic driving force.

The chemical potential of a solution that is in equilibrium with a cluster of size $n$ at the surface is given by
\begin{equation}
\mu\left(n\right)=k_{B}T\ln\left(c\left(n\right)/c_{0}\right)
\label{eq:s7}
\end{equation}
using the concentration of the solution $c\left(n\right)$ at the surface of cluster because $c\left(n\rightarrow\infty\right)=c_{\rm sat}$~\cite{Peter2011} and $\mu\left(n\rightarrow\infty\right)=-\Delta\mu_{0}$ from Eq.~(\ref{eq:s6}).  Equations.~(\ref{eq:s6}) and (\ref{eq:s7}) lead to the well-known Ostwald-Freundlich equation~\cite{Peter2011}
\begin{equation}
c\left(R\right)=c_{0}\exp\left[-\beta\Delta\mu_{0} R_{*}\left(\frac{1}{R_{*}}-\frac{1}{R}\right)\right]
\label{eq:s8}
\end{equation}
where 
\begin{equation}
R_{*} = \frac{2\gamma v_{\rm m}}{\Delta\mu_{0}}
\label{eq:s10}
\end{equation}
is the radius of critical nucleus with its free energy that corresponds to the energy barrier of nucleation given by 
\begin{equation}
\Delta G_{*}=\frac{16\pi\gamma^{3}v_{\rm}^{2}}{3\Delta\mu_{0}^{2}}
                =\frac{4\pi\gamma}{3}R_{*}^{2}
\label{eq:s10x}
\end{equation}              
determined {\it thermodynamically} from $\partial \Delta G/\partial n = 0$ in Eq.~(\ref{eq:s3}), and $v_{\rm m}$ is the molar volume defined as $v_{\rm m}n=\left(4\pi/3\right)R^{3}$ and is related to $\phi$ in Eq.~(\ref{eq:s3}) as $v_{m}=\phi^{3/2}/\left(3\left(4\pi\right)^{1/2}\right)$.  In Eq.~(\ref{eq:s8}) we have expressed the concentration at the surface, $c(R)$, as a function of the radius, $R$, instead of the number of monomers, $n$.  The boundary condition (\ref{eq:s8}) assumes that the chemical equilibrium is reached so quickly as the radius of cluster growths. This equation is also known to
represent the Gibbs-Thomson effect~\cite{Saito1996} which is directly observable even on solid surface~\cite{McLean1997}. 

From Eq.~(\ref{eq:s7}) we have 
\begin{equation}
\mu\left(n\right)=-\Delta\mu_{0} R_{*}\left(\frac{1}{R_{*}}-\frac{1}{R}\right).
\label{eq:s9}
\end{equation}
By replacing $\mu\left(n\right)$ in Eq.~(\ref{eq:s5}) with Eq.~(\ref{eq:s8}), we can obtain the well-known formula~\cite{Shneidman2001} 
\begin{equation}
\frac{dR}{dt}=\frac{v_{m}\beta w\left(n\right)\Delta\mu_{0} R_{*}}{4\pi R^{2}}\left(\frac{1}{R_{*}}-\frac{1}{R}\right)
\label{eq:s11}
\end{equation}
that can be used for determining cluster growth~\cite{Shneidman2001} given the rate of attachment, $w\left(n\right)$.

Slezov~\cite{Slezov2009} showed that the rate of attachment, $w\left(n\right)$, for the post-rcritical nucleus could be determined in the case the post-critical nucleus grows by diffusion and material attachment~\cite{Reiss1951,Grinin2004}, by considering the Zeldovich equation (\ref{eq:s11}) as the diffusional growth equation as follows. 
\begin{equation}
\frac{dR}{dt} = v_{m} j_{R}
\label{eq:s12}
\end{equation}
where
\begin{equation}
j_{R} = D\left.\left(\frac{\partial c}{\partial r}\right)\right|_{r=R}
\label{eq:s13}
\end{equation}
is the diffusion flux at the surface of growing droplet, where $D$ is the solute diffusivity.  The concentration field $c\left(r,t\right)$ obeys the diffusion equation~\cite{Grinin2004}
\begin{equation}
\frac{\partial c}{\partial t}=\frac{D}{r}\frac{\partial^{2}}{\partial r^{2}}\left(r c\left(r,t\right)\right)
\label{eq:s14}
\end{equation}
for spherical symmetry, the steady state ($t\rightarrow\infty$) solution of which is given by
\begin{equation}
c(r) = c_{0}-\left(c_{0}-c\left(R\right)\right)\frac{R}{r}.
\label{eq:s15}
\end{equation}
Then, the diffusion flux is simply given by
\begin{equation}
j_{R}=D\frac{c_{0}-c(R)}{R},
\label{eq:s16}
\end{equation} 
and Eq.~(\ref{eq:s12}) is given by
\begin{equation}
\frac{dR}{dt}=\frac{v_{m} D c_{0}}{R}\left(1-\frac{c\left(R\right)}{c_{0}}\right),
\label{eq:s17}
\end{equation}
which can be written as
\begin{equation}
\frac{dR}{dt}=v_{m} D\beta c_{0} \Delta\mu_{0}\frac{R_{*}}{R}\left(\frac{1}{R_{*}}-\frac{1}{R}\right),
\label{eq:s18}
\end{equation}
by expanding $c(R)$ in Eq.~(\ref{eq:s8}) around $R_{*}$.  Therefore, the diffusional growth equation can be written in the form
\begin{equation}
\frac{dR}{dt} \propto \left(\frac{1}{R_{k}}-\frac{1}{R}\right),
\label{eq:s19}
\end{equation}
which is similar to the Zeldovich equation (\ref{eq:s11}), in which the {\it kinetic} critical radius $R_{k}$ coincides with the {\it thermodynamic} critical radius $R_{*}$ defined by Eq.~(\ref{eq:s10}). The rate of attachment in Eq.~(\ref{eq:s11}) is now given by
\begin{equation}
w\left(n\right)=4\pi R D c_{0}
\label{eq:s20}
\end{equation}
for the post-critical nucleus~\cite{Slezov2009}.  We should note in passing that the specific form Eq.~(\ref{eq:s18}) is {\it not} directly related to the Ostwald-Freundlich equation given by Eq.~(\ref{eq:s8}), but is related indirectly through the specific form of the steady state diffusion flux given by Eq.~(\ref{eq:s16}). 

\begin{figure}[htbp]
\begin{center}
\includegraphics[width=0.8\linewidth]{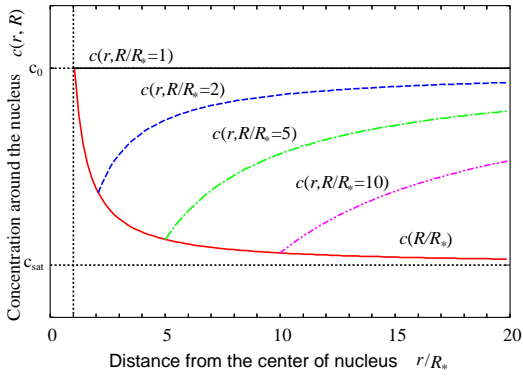}
\caption{
Concentrations $c(r)=c\left(r, R/R_{*}\right)$ of monomers around various post-critical nucleus having radius $R/R_{*}=1, 2, 5, 10$ as a function of distance $r$ from center of nucleus and concentration $c(R)$ at surface of growing nucleus, as determined from Ostwald-Freundlich boundary condition . The concentration around the critical nucleus ($R/R_{*}=1$) is a constant $c\left(r\right)=c_{0}$.  As the nucleus grows as the post-critical nucleus, the concentration at the surface of nucleus starts to decrease according to the Ostwald-Freundlich equation. Then, a concentration gradient appears around the nucleus, according to Eq.~(\ref{eq:s15}), and the diffusional flux starts to flow in and feeds the growing nucleus. }
\label{fig:1}
\end{center}
\end{figure}

Therefore, as soon as the embryo crosses the thermodynamic saddle point at $R=R_{*}$ as the critical nucleus, the solution concentration $c(R)$ at the surface of the nucleus with radius $R>R_{*}$ starts to decrease (Fig..~\ref{fig:1}), and the concentration around the supercritical nucleus starts to decrease (Fig.~\ref{fig:1}), which in turn, induces a diffusion flux according to Eq.~(\ref{eq:s13}).  Then, the nucleus continue to grow according to Eq.~(\ref{eq:s18}) as if it just crossed the kinetic critical point at $R=R_{k}=R_{*}$.   Therefore, the critical radius, $R_{*}$, in Eq.~(\ref{eq:s11}) should be interpreted as the thermodynamic critical radius fixed by the Ostwald-Freundlich boundary condition, whereas the $R_{*}$ in the kinetic equation Eq.~(\ref{eq:s18}) should  be interpreted as the kinetic critical radius determined from the diffusion flux.  In Fig.~\ref{fig:1}, we can observe that the size of the depletion zone or diffusion shell around the nucleus increases as the size of the nucleus increases.  In fact, the self-similar theory~\cite{Adzhemyan2006} predicts that the size of the depletion zone scales with the size of a droplet.  Furthermore, the size depends on the supersaturation~\cite{Kuchma2009}, and Fig.~\ref{fig:1} corresponds to weak supersaturation, where the depletion zone is wider than the nucleus radius.

Thus far, we have assumed that the diffusion is very fast so that not only does the concentration at the surface of the nucleus quickly reach equilibrium as defined by the Ostwald-Freundlich boundary condition (\ref{eq:s8}) but the concentration field and flux also reach the steady state defined by Eqs.~(\ref{eq:s15}).   In fact, there must be a time delay in reaching the steady state, and the time-dependent concentration field is approximately given by~\cite{Kuni2003,Grinin2004} 
\begin{equation}
c(r) = c_{0}-\left(c_{0}-c\left(R\right)\right)\frac{R}{r}.\left[1-\Phi\left(\frac{r-R}{2\sqrt{D t}}\right)\right]
\label{eq:s21}
\end{equation}
with
\begin{equation}
\Phi\left(u\right)=\frac{2}{\sqrt{\pi}}\int^{u}_{0}\exp\left(-\xi^{2}\right)d\xi
\label{eq:s22}
\end{equation}
instead of Eq.~(\ref{eq:s15}).  Then, the growth law given by Eqs.~(\ref{eq:s16}) and (\ref{eq:s18}) is modified as~\cite{Kuni2003}
\begin{equation}
\frac{dR}{dt}=v_{m} D\beta c_{0} \Delta\mu_{0}\frac{R_{*}}{R}\left(\frac{1}{R_{*}}-\frac{1}{R}\right)\left(1+\frac{R}{\sqrt{Dt}}\right),
\label{eq:s23}
\end{equation}
so that the resulting growth velocity at the beginning of growth is faster than that predicted from the steady state flux obtained using Eq.~(\ref{eq:s18}).  However, the kinetic critical radius remains the same as the thermodynamic critical radius ($R_{k}=R_{*}$).

As has been noted below Eq.~(\ref{eq:s20}), the specific form Eq.~(\ref{eq:s18}) is not ascribed to the Ostwald-Freundlich boundary condition but solely to the diffusion flux given by Eq.~(\ref{eq:s16}).  Any subtle change in the nucleation flux given by Eq.~(\ref{eq:s13}) may lead to a kinetic critical radius $R_{k}$ defined using the diffusional growth equation (\ref{eq:s19}), which would be different from the thermodynamic critical radius $R_{*}$ defined using Eq.~(\ref{eq:s11}).  For example, a temporal depletion or superabundance of materials in the mother phase leads to the flux
\begin{equation}
j_{R}=D\frac{c_{1}-c(R)}{R}
\label{eq:s24}
\end{equation}
instead of Eq.~(\ref{eq:s16}) with $c_{1}>c_{0}$ for superabundance and $c_{1}<c_{0}$ for depletion.  Then, the  growth equation has the same form as that of Eq.~(\ref{eq:s19}) with
\begin{equation}
\frac{1}{R_{k}}=\frac{1}{R_{*}}+\frac{c_{1}}{c_{0}\beta\Delta\mu_{0}R_{*}}\left(1-\frac{c_{0}}{c_{1}}\right)
\label{eq:s25}
\end{equation}
When the mother phase is depleted ($c_{1}<c_{0}$), the kinetic critical radius, $R_{k}$, becomes larger than the thermodynamic critical radius, $R_{*}$ ($R_{k}>R_{*}$), from Eq.~(\ref{eq:s25}).  Then, the {\it thermodynamic} post-critical nucleus turns to the {\it kinetic} pre-critical nucleus, and it shrinks kinetically~\cite{Peter2011}.   In contrast, the kinetic critical radius is smaller than the thermodynamic critical radius ($R_{k}<R_{*}$) when the mother phase is superabundant ($c_{1}>c_{0}$).  Then, the thermodynamically pre-critical nucleus turns to a post-critical nucleus, and it grows kinetically.  Thereafter, the nucleation rate is enhanced.  Such enhanced nucleation in a concentrated solution is typically found~\cite{Erdemir2009, Sear2012}, for example, in the crystallization of proteins~\cite{Vekilov2012} and small molecules~\cite{Bonnett2003}. 

 It is well known that a rapid decrease of supersaturation $\Delta \mu_{0}$ due to the depletion of monomer $c_{0}\rightarrow c_{1}<c_{0}$ causes a drastic effect to the growth of nucleus~\cite{LaMer1950,Gorshkov2010,Nanev2011}. Our simple model has predicted that the thermodynamic critical nucleus turns to that of the kinetic pre-critical nucleus from Eq.~(\ref{eq:s25}) and the nucleation will be hindered. Also, the thermodynamic critical radius $R_{*}$ in Eq.~(\ref{eq:s10}) and the thermodynamic energy barrier $\Delta G_{*}$ in Eq.~(\ref{eq:s10x}) increases as the chemical potential $\Delta\mu_{0}$ in Eq.~(\ref{eq:s4}) decreases due to the depletion of monomer. In fact, the increase of the thermodynamic critical radius defined by Eq.~(\ref{eq:s10}) is given by the same formula Eq.~(\ref{eq:s25}).  Therefore, the nucleation becomes less probable not only kinetically but also thermodynamically and will be hindered in later time~\cite{Gorshkov2010,Nanev2011}. Our kinetic picture is consistent to this scenario.  Moreover, our discussion has clearly indicated that it can happen even temporally and locally due to the temporal fluctuation of diffusion flux by the depletion of local concentration. 

Our discussion, however, has concentrated on the stage where the critical nucleus just turns to the growing nucleus by diffusion. Therefore, subsequent Ostwald ripening~\cite{Slezov2009,Iwamatsu1999} of single nucleus or the aggregation of multiple nuclei~\cite{Gorshkov2010} during the growth stage is beyond the scope of the present study. Also, since we have considered only the material diffusion and neglected the heat flow, the instability and fractal growth~\cite{Mullins1963} of nucleus cannot be discussed within our present formalism.

In conclusion, we showed semi-analytically that any subtle change in the diffusion flux leads to a change in the kinetic critical radius, $R_{k}$, from the thermodynamic critical radius, $R_{*}$, even when Ostwald-Freundlich boundary condition is maintained.  The same conclusion has already been reached by Peter~\cite{Peter2011} using numerical simulation.  Incidentally, a recent development in self-similar solutions~\cite{Adzhemyan2006,Grinin2011} to diffusional growth assumes a large $R$ limit for the growth equation Eq.~(\ref{eq:s18}) and is unsuitable for discussing nucleus growth near the critical radius, $R_{*}$.

\begin{acknowledgments}
This work was supported under a project for strategic advancement of research infrastructure for private universities, 2009-2013, operated by MEXT, Japan. 
\end{acknowledgments}


\begin{thebibliography}{99}
\bibitem{Kelton2010} K. F. Kelton and A. L. Greer, Nucleation in Condensed Matter, Applications in Materials and Biology, Pergamon, Oxford, 2010, Chapter 6.
\bibitem{Gorshkov2010} V. Gorshkov and V. Privman, Physica E {\bf 43}, 1 (2010).
\bibitem{Meldrum2008} F. C. Meldrum and H. C\"ofen, Chem. Rev. {\bf 108}, 4332 (2008).
\bibitem{Sear2012} R. P. Sear, Int. Mat. Rev. {\bf 57}, 328 (2012).
\bibitem{Zener1949} C. Zener, J. Appl. Phys. {\bf 20}, 950 (1949).
\bibitem{Frank1950} F. C. Frank, Proc. R. Soc. Lond. A {\bf 201}, 586 (1950).
\bibitem{Epstein1950} P. S. Epstein and M. S. Plesset, J. Appl. Phys. {\bf 18}, 1505 (1950).
\bibitem{LaMer1950} V. K. LaMer and R. H. Dinegar, J. Am. Chem. Soc. {\bf 72}, 4847 (1950).
\bibitem{Reiss1951} H. Reiss, J. Chem. Phys. {\bf 19}, 482 (1951).
\bibitem{Privman1999} V. Privman, D. V. Goia, J. Park, and E. Matijevi\'c, J. Colloid Interface Sci. {\bf 213}, 36 (1999).
\bibitem{Robb2008} D. T. Robb and V. Privman, Langmuir {\bf 24}, 26 (2008).
\bibitem{Kuchma2009a} A. E. Kuchma, G. Yu. Gor, and F. M. Kuni, Colloid J {\bf 71}, 520 (2009).
\bibitem{Grinin2011} A. P. Grinin, G. Yu. Gor, F. M. Kuni, Atoms. Res. {\bf 101}, 503 (2011).
\bibitem{Kalikmanov2013} K. I. Kalikmanov, Nucleation Theory, Springer, Heidelberg, 2013.
\bibitem{Wilemski1995} G. Wilemski and B. E. Wyslouzil, J. Chem. Phys. {\bf 103}, 1127 (1995).
\bibitem{Wyslouzil1995} B. E. Wyslouzil and G. Wilemski, J. Chem. Phys. {\bf 103}, 1137 (1995). 
\bibitem{Kathmann2004} S. M. Kathmann, G. K. Schenter, and B. C. Garrett, J Chem. Phys. {\bf 120}, 9133 (2004).
\bibitem{Zeldovich1943} Ya. B. Zeldovich, Acta Physicochim URSS {\bf 18}, 1 (1943).
\bibitem{Slezov2009} V. V. Slezov, Kinetics of First-order Phase Transition, Wiley-VCH, Weinheim, 2009, Chapter 5.

\bibitem{Iwamatsu2012} M. Iwamatsu, J. Chem. Phys. {\bf 136}, 204702 (2012).
\bibitem{Slezov1998} V. V. Slezov and J. W. P. Schmelzer, J. Phys. Chem. Sol. {\bf 59}, 1507 (1998).
\bibitem{Peter2011} B. Peter, J. Chem. Phys. {\bf 135}, 044107 (2011).
\bibitem{Saito1996} Y. Saito, Statistical Physics of Crystal Growth, World Scientific, Singapore 1996, Part. III. 
\bibitem{McLean1997} J. G. McLean, B. Krishnamachari, D. R. Peale, E. Chason, J. P. Sethna and B. H. Cooper, Phys. Rev. B {\bf 55}, 1811 (1997).     

\bibitem{Shneidman2001} V. A. Shneidman, J. Chem. Phys. {\bf 115}, 8141 (2001).
\bibitem{Grinin2004} A. P. Grinin, A. K. Shchekin, F. M. Kuni, E. A. Grinina and H. Reiss, J. Chem. Phys. {\bf 121}, 387 (2004).


\bibitem{Adzhemyan2006} L. Ts. Adzhemyan, A. N. Vasil'ev, A. P. Grinin, and A. K. Kazansky, Colloid J {\bf 68}, 381 (2006).
\bibitem{Kuchma2009} A. E. Kuchma, F. M. Kuni, and A. K. Shchekin, Phys. Rev. E {\bf 80}, 061125 (2009).
\bibitem{Kuni2003} F. M. Kuni, E. A. Grinina, and A. K. Shchekin, Colloid J {\bf 65}, 809 (2003).

\bibitem{Erdemir2009} D. Erdemir, A. Y. Lee, and A. S. Myerson, Acc. Chem. Res. {\bf 42}, 621 (2009).
\bibitem{Vekilov2012} P. G. Vekilov, J. Phys.: Condens Matter {\bf 24}, 193101 (2012).
\bibitem{Bonnett2003} P. E. Bonnett, K. J. Carpenter, S. Dawson and R. J. Davery, Chem. Commun. {\bf 6}, 698 (2003).
\bibitem{Nanev2011} C. N. Nanev, F. V. Hodzhagolu, and I. L. Dimitrov, Cryst. Growth Des. {\bf 11}, 196 (2011).  
\bibitem{Iwamatsu1999} M. Iwamatsu, J. Appl. Phys. {\bf 86}, 5541 (1999). 
\bibitem{Mullins1963} W. W. Mullins and R. F. Sekerka, J. Appl. Phys. {\bf 34}, 323 (1963).  
\end{thebibliography}

\end{document}